\documentstyle[12pt,dina4,epsf,psfig]{article}
\titlepage

\begin{document}
\title{Phasespace Correlations of
Antideuterons in  \\
Heavy Ion Collisions\footnote{Work supported by Bundesministerium f\"ur
Forschung und Technik, Deutsche Forschungsgemeinschaft, Gesellschaft
f\"ur Schwerionenforschung}
}
\author{M.~Bleicher, C.~Spieles, A.~Jahns, R.~Mattiello,\\
H.~Sorge, H.~St\"ocker and W.~Greiner
\\ \\
Institut f\"ur Theoretische Physik \\
Johann Wolfgang Goethe Universit\"at \\ Frankfurt am Main, Germany }
\date{}
\maketitle
\vspace*{1.0cm}
\begin{abstract}
\noindent
In the framework of the relativistic quantum molecular dynamics
approach ({\small RQMD}) we investigate antideuteron ($\overline{d}$)
observables in Au+Au collisions at 10.7~AGeV.
The impact parameter dependence of the formation ratios
$\overline{d}/\overline{p}^2$
and ${d}/{p}^2$ is calculated. In central collisions, the
antideuteron formation ratio is predicted to be two orders of magnitude
lower than the deuteron formation ratio. The $\overline{d}$ yield in
central Au+Au collisions is one order of magnitude lower than in Si+Al
collisions. In semicentral collisions different
configuration space distributions of $\overline{p}$'s and $\overline{d}$'s
lead to a large ``squeeze--out'' effect for antideuterons, which is not
predicted for the $\overline{p}$'s.
\end{abstract}


\newpage
\noindent
To understand recent measurements of the antideuteron
formation ratio (multiplicity of $\overline{d}$ divided by the
$\overline{p}$ multiplicity squared, both at the same momentum per nucleon)
in Si+Au collisions at the {\small AGS}, it has been suggested that
the shape and size of the antinucleon source, which is different from
the nucleon source, has to be taken into account \cite{mrc93,aok92,jna94}.
Enhanced production of antibaryons and antimatter clusters has been
proposed as potential signature of a quark--gluon
plasma \cite{heinz,rischke90}. The yield of light (anti--)nuclei
is connected to baryon density \cite{gavin} and its spatial distribution
in the later
stages of the collision. This can be used to extract source sizes in an
alternative way to the {\small HBT}--analysis \cite{jna94b}.
\\
Production processes which enhance the $\overline{p}$ yield can be
counter--balanced by the annihilation in the baryon rich
environment \cite{gavin,aja94}. This should be observable via the
``anti--flow'' correlation between matter and antimatter \cite{aja94}.
\\
Strong annihilation distorts the
configuration space distribution of antibaryons and strongly affects the
antideuteron production as shown below. The strong suppression of antimatter
cluster formation and their characteristic phasespace distribution of
antideuterons for finite impact parameters is discussed. These phenomena
can be experimentally scrutinized to differentiate
between annihilation assumptions. It can also reveal details of the
structure of the eventshape unaccessible by $\overline{p}$ or p data.
\\
The calculations presented here are based on a microscopic phase
space approach, the relativistic quantum molecular dynamics model
({\small RQMD} 1.07--cascade mode)\cite{RQMD1}.
This model is based on the propagation of all hadrons on classical
trajectories in the framework of Hamilton constraint dynamics.
{\small RQMD} includes secondary interactions, e.g. the annihilation of
produced mesons on baryons, which may lead to the formation of $s$ channel
resonances or strings. The absorption probability for $\bar p$'s in the
baryonic medium is determined by the free $p\overline{p}$ annihilation
cross section.
\\
Since {\small RQMD} does not include the production of light
(anti--)nuclei dynamically, cluster formation is added after strong
freeze--out.
We calculate the deuteron and antideuteron formation probability by
projecting the
(anti--)nucleon pair phasespace on the (anti--)deuteron wave function via
the Wigner--function method as described in \cite{rma94,gyu83}.
The Hulth\'en
parametrization \cite{hulthen} is used to describe the relative part of the
(anti--)deuteron wavefunction.
The yield of antideuterons (and deuterons, resp.) is given by
\\
\centerline{${\rm d}N_{\overline{d}}={1\over 2}{3\over
4}\Big<\sum_{i,j}\rho^{^{\rm W}}_{\overline {d}}
(\Delta \vec{R},\Delta \vec{P})\Big>
{\rm d}^3(p_{i_{\overline{p}}}+p_{j_{\overline{n}}})$.}
\\
The sum goes over all $\overline{n}$ and $\overline{p}$
pairs, whose relative distance ($\Delta \vec{R}$) and relative momentum
($\Delta \vec{P}$) are calculated in their rest frame at the earliest
time after both nucleons have ceased to interact. The
factors ${1\over 2}$ and ${3\over 4}$ account for the statistical spin and
isospin projection on the (anti--)deuteron state. The calculation of the
$\bar t$ is straight forward, by exchanging the Hulth\'en
parametrization of the $\bar d$ wavefunction with a 3--body harmonic oscilator
wavefunction to describe the $\bar t$ \cite{rma94}.
We use the event mixing technique to improve statistics.
\\
It has been shown that the {\small {\small RQMD}/C}--Model describes the
Si(14.6AGeV)+Au antideuteron data reasonably well \cite{jna94}.
\\
The configuration space distribution of midrapidity antideuterons is
calculated
at freezeout in a  cut perpendicular to the beam axis ($\Delta z=\pm 0.5$fm)
for the Au(10.7 AGeV)+Au, b=5~fm reaction (fig.~1). Antimatter is
preferentially emitted from the surface of the fireball -- i.e.
the region of hot and dense
matter -- in clear contrast to the source of baryons, which spreads over the
whole reaction volume. This can be understood as a consequence of antibaryon
absorption, which produces deep holes in the momentum-- (at
$p_{\rm cms}\approx 0$~GeV/c) and even more
so in the configuration space distribution (at $x_{\rm cms}\approx
0$~fm, $y_{\rm cms}\approx 0$~fm, fig.~1)
of the antibaryons at freezeout \cite{aja94,mrc93}.
\\
The antideuteron formation happens perpendicular to the reaction plane, where
the observer gets an undistracted view directly into the hot and dense
participant matter. The suppression of $\overline{p}$'s in the
reaction plane is due to the fact that the baryons annihilate
$\overline{p}$'s preferentially moving parallel to the impact parameter plane.
The anticluster formation probability decreases rapidly with the
relative distance of the $\overline{p}$, $\overline{n}$ constituents.
The small decrease of the spatial
$\overline{p}$ density ($<$30\%) along the football--shaped region in the
$x$-$y$--plane shows up very clearly in the $\overline{d}$'s.
\\
The anisotropic configuration space distribution of the
$\overline{d}$'s can be magnified via the azimuthal ($\phi$) distribution of
antideuterons. $\phi$ is the angle between $\vec{p}_{_T}$ and the $x$--axis
($\tan \phi = |p_y | /p_x $). Thus, a vector with $\phi =0$ degrees
points into the direction of the $x$--axis.
\\
Fig.~2 shows the azimuthal distribution of midrapidity
antiprotons (full line), antideuterons (long dashed) and
antitritons (short dashed)
in peripheral Au(10.7AGeV)+Au collisions. The momentum
distribution of $\overline{p}$'s in the $p_x$-$p_y$--plane is slightly
concave and reflects the geometry of the almond shaped reaction zone.
\\
In contrast, the $\overline{d}$ distribution is
shifted towards $\phi=90^{\circ}$, in line with the
described configuration space distribution of the $\overline{d}$'s.
This looks like a ''squeeze--out'' effect. It is expected that the effect is
even
more pronounced for $\overline{t}$'s.
The asymmetry in the azimuthal angular distributions
(${\rm d}^2N_{\overline{d}}/{\rm d}\phi/{\rm d}y|_{\rm y=y_{mid}}$) can
be experimentally tested -- once the reaction plane is determined.
The exact reflection symmetry of fig.~2 is due to the fact that every
event for the symmetric system has been reflected by $\phi =90^{\circ}$.
\\
Let us explore the reaction volume dependence of the
(anti--)deuteron formation ratio. This can be done by varying the impact
parameter b as shown
in fig.~3 ($\overline{d}/\overline{p}^2$--ratio, full circles; $d/p^2$--ratio,
open squares). In thermodynamic approaches the (anti--)deuteron formation
rate is proportional to the inverse volume of the sources. Those rates
measure the average phase--space distance of nucleons and antinucleons,
respectively \cite{aja94}.
\\
The $\overline{d}$
formation ratio in central collisions (where up to $95\%$ of the
initially produced antinucleons are reabsorbed) is predicted to be
roughly two orders of magnitude
lower than the formation ratio of deuterons. This difference vanishes
when going to higher impact parameters or small systems (like Si+Al).
It has to be pointed out that this effect is so strong that the absolute yield
of $\overline{d}$ actually {\it decreases} from $\approx 10^{-7}$
$\overline{d}$'s per Si+Al event
to $\approx 10^{-8}$ $\overline{d}$'s in central Au+Au collisions!
If the antideuteron
formation ratio is calculated within a momentum coalescence model with a
momentum cutoff parameter of $\Delta p = 120$~MeV
($\overline{d}/\overline{p}^2$--ratio, open circles), one finds only little
sensitivity to the impact parameter chosen. This strengthens the explanation of
the mostly configuration space origin of this effect. For very peripheral
collisions the antideuteron formation ratio must decrease rapidly
(not shown in the figure), because the production
of two antibaryons is not allowed due to the limited available energy.
The E814 collaboration might be able to measure these dependencies as they have
done for matter clusters \cite{E814}.
\\
The predicted reaction volume dependence and the squeeze--out of
anti--fragments reflects mainly the spatial distribution of antibaryons.
Antimatter clusters provide an exciting opportunity to
measure the spatial properties of the exploding dense matter.

\newpage
\vspace{1.0cm}

\newpage

{\LARGE Figure Captions : }

\vspace{1.5cm}

{\noindent \Large Figure 1 } \\
Configuration space distribution of midrapidity antideuterons
in the reaction Au(10.7~AGeV)+Au, b=5fm at freezeout in a cut
perpendicular to the reaction plane. The initial shapes of the colliding nuclei
are also plotted.

\vspace{1cm}

{\noindent \Large Figure 2 } \\
Correlations of antiprotons lead to a ``squeeze--out'' for $\overline d$
and $\overline t$ in peripheral Au$(10.7$AGeV$)$+Au reactions
due to a lower anticluster formation probability in the reaction plane.

\vspace{1cm}

{\noindent \Large Figure 3 } \\
Impact parameter dependence of the (anti--)deuteron to the
(anti--)proton ratio squared ($\overline{d}/\overline{p}^2$,full
circles; $d/p^2$, open squares).
Open circles: the result of a simple momentum coalescence model.
The lines are to the eye.

\newpage
\vspace{2cm}
\noindent
{\Large Figure 1}
\psfig{figure=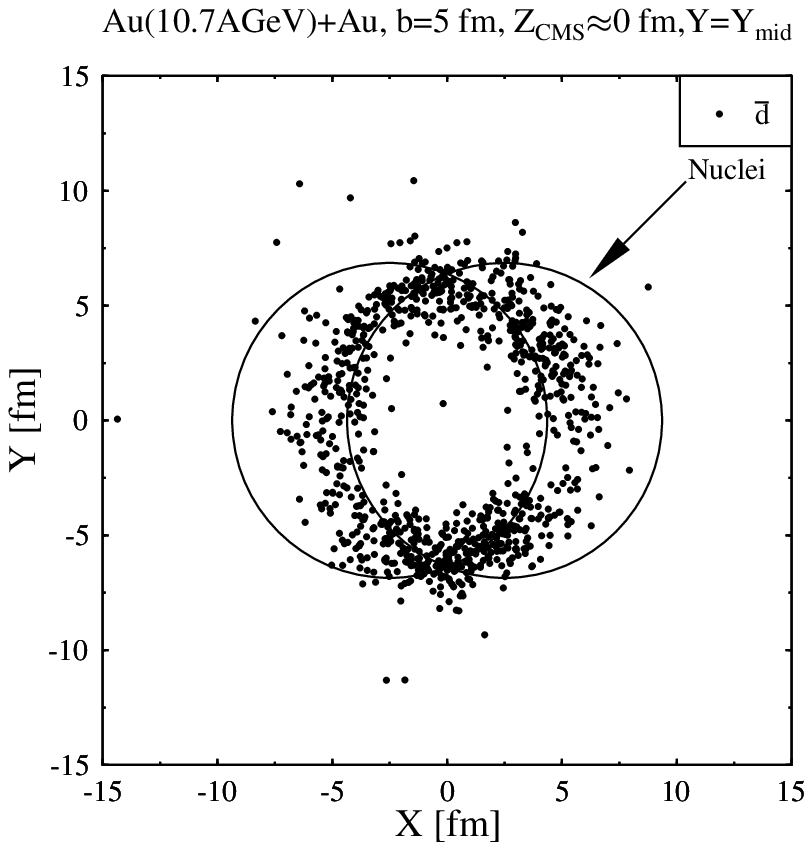,width=13cm}

\newpage
\vspace{2cm}
\noindent
{\Large Figure 2}
\psfig{figure=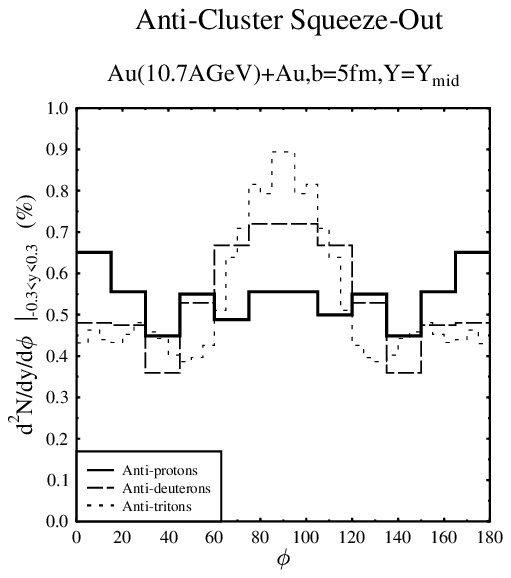,width=13cm}

\newpage
\vspace{2cm}
\noindent
{\Large Figure 3}
\psfig{figure=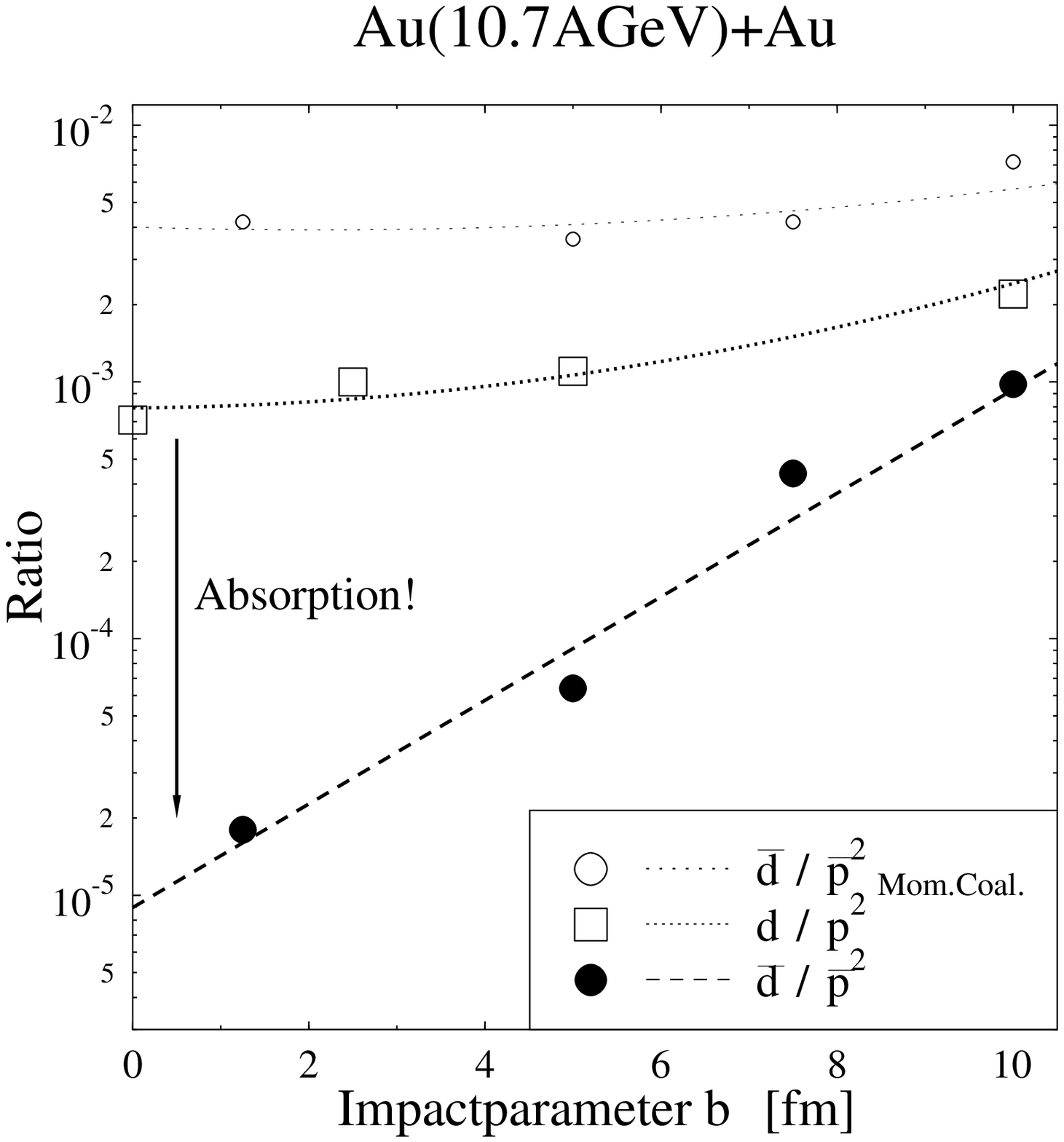,width=13cm}


\begin{thebibliography}{mmmmm}
%
\bibitem{mrc93}
St. Mrowczynski: Phys. Lett. B308 (1993)216
\bibitem{aok92}
A. Aoki {\it et al.}: Phys. Rev. Lett. 69 (1992)2345
\bibitem{jna94} J.L. Nagle, B.S. Kumar, M.J. Bennet, S.D. Coe, G.E.
Diebold, J.K. Pope, A. Jahns, H. Sorge: Phys. Rev. Lett. 73 (1994)2417
\bibitem{heinz} U. Heinz, P.R. Subramanian, H. St\"ocker, W. Greiner:
J. Phys. G: Nucl. Phys. 12 (1986)1237
\bibitem{rischke90}
D. H. Rischke {\it et al.}: Phys. Rev. D41 (1990)111
\bibitem{gavin}
S. Gavin, M. Gyulassy, M. Pl\"umer, R. Venugoplan: Phys. Lett. B234
(1990)175
\bibitem{jna94b} J.L. Nagle, B.S. Kumar, M.J. Bennet, G.E.Diebold,
J.K. Pope, H. Sorge, J.P. Sullivan: Phys. Rev. Lett. 73 (1994)1219
\bibitem{aja94}
A. Jahns, C. Spieles, H. Sorge, H. St\"ocker, W. Greiner: Phys. Rev.
Lett. 22 (1994);
A. Jahns, C. Spieles, R.Mattiello, H. Sorge, H. St\"ocker,
W. Greiner: Phys. Lett. B308 (1994)3464
\bibitem{RQMD1}
H. Sorge, H. St\"ocker, W. Greiner: Ann. Phys. (N.Y.) 192 (1989) 266;
\bibitem{rma94}
R. Mattiello, A. Jahns, H. Sorge, H. St\"ocker, W. Greiner: Phys. Rev.
Lett. 74 (1995)2180
\bibitem{gyu83}
M. Gyulassy, K. Frankel, E.A. Remler: Nucl. Phys. A402 (1983)596
\bibitem{hulthen} L. Hulthen: Ark. Mat. Ast. Fys. 28, No.5
\bibitem{E814} J. Barrette {\it et al.} (E814 collab):
Phys. Rev. C50 (1994)1077
\end{thebibliography}
\end{document}